\documentclass[aps,prl,twocolumn,superscriptaddress]{revtex4}
\usepackage{graphicx}
\usepackage{amssymb}
\usepackage{amsmath}
\usepackage{graphicx}
\usepackage{epsfig}
\usepackage{color}

\begin{document}

\title{Scaling law of the Hall coefficient in cuprates}
\author{Yi-feng Yang}
\email[]{yifeng@iphy.ac.cn}
\affiliation{Beijing National Laboratory for Condensed Matter Physics and Institute of Physics,
Chinese Academy of Sciences, Beijing 100190, China}
\affiliation{School of Physical Sciences, University of Chinese Academy of Sciences,
Beijing 100190, China}
\affiliation{Songshan Lake Materials Laboratory, Dongguan, Guangdong 523808, China}
\date{\today }

\begin{abstract}
One parameter scaling of the Hall coefficient of cuprates has been well known for almost three decades, but still lacks a simple mathematical expression. Motivated by the recent phenomenological prediction of the universal scaling of the thermal Hall conductivity, we propose here a simple scaling function for the Hall coefficient in cuprates that varies exponentially with the temperature. Comparison with experimental data in La$_{2-x}$Sr$_{x}$CuO$_{4}$, YBa$_2$Cu$_3$O$_{6+\delta}$, and Bi$_{2}$Sr$_{2-x}$La$_{x}$CuO$_{6+\delta }$ confirms its validity over a wide temperature and doping range. The scaling is independent of microscopic details and arises as long as the hole carriers have nonzero Berry curvature density within a finite energy window, possibly due to their interaction with the spin liquid. This differs from the activation model proposed previously to explain the Hall coefficient at low doping and high temperatures, and suggests a unified picture in line with the one parameter scaling observed in many physical quantities of cuprates.
\end{abstract}

\maketitle

The existence of topological excitations has long been speculated in cuprates but still remains unsettled \cite{Anderson1987Science,Lee2006RMP}. Finding experimental evidences or even clues is notoriously difficult. A recent advance is the discovery of large thermal Hall signals in both undoped and underdoped samples \cite{Grissonnanche2019Nature}. Later analysis reveals a universal contribution from charge neutral carriers with nonzero Berry curvature density, which decreases exponentially with increasing temperature over a wide intermediate temperature range \cite{Yang2020PRL}. Although the true mechanism of the thermal Hall effect has not been decided, it probably reflects the presence of some exotic excitations of the spin liquid \cite{Han2019PRB} or chiral phonons with a finite Berry curvature density induced by coupling to the spins \cite{Grissonnanche2020NP}. In the best scenario, some intrinsic topological properties may exist and even have influence beyond the thermal Hall effect.

In this work, we explore this idea and show that a similar scaling law also exists in the Hall coefficient and explains the behavior of the Hall angle, suggesting that doped holes may also inherit some topological properties possibly by interacting with the spin liquid. Anderson has proposed that the cotangent of the Hall angle should satisfy a $T^2$ law due to the spin-charge separation, but it generally fails in underdoped and overdoped regions \cite{Anderson1991PRL}. The Hall coefficient has also been interpreted based on an activation model \cite{Gorkov2006PRL,Ono2007PRB}, in which charge carriers are thermally activated across the charge transfer gap and their normal Hall coefficient was found to give a good fit of the experimental data in La$_{2-x}$Sr$_{x}$CuO$_{4}$ at low doping and high temperatures. However, the fitting fails below 300 K or beyond optimal doping ($x=0.15$) where doped holes become dominant \cite{Ono2007PRB}. Though probably useful for explaining the Hall data in La$_{2}$CuO$_{4}$, its methodology is against the observation of one parameter scaling of the Hall coefficient that has been well established over a wide doping range for almost three decades \cite{Hwang1994PRL,Luo2008PRB}. Our work provides an alternate mechanism with a simple scaling function that not only respects the one parameter scaling, but covers a much wider temperature and doping range. It also connects the Hall transport of charge carriers and charge neutral excitations and hints a unified picture of the basic cuprate physics.

\begin{figure}[t]
\centering\includegraphics[width=0.5\textwidth]{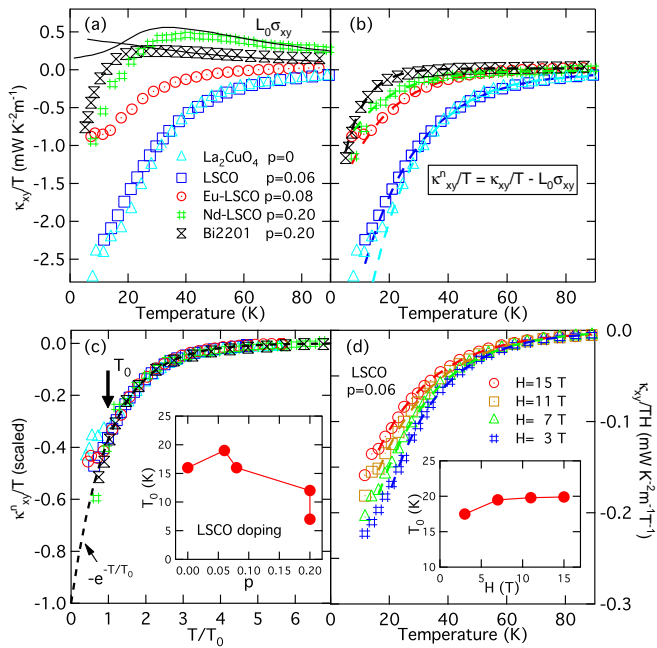}
\caption {(a) Thermal Hall conductivity of five cuprate samples below the critical doping, reproduced from experiment \cite{Grissonnanche2019Nature}. The solid lines denote the contribution of charge carriers estimated from the Wiedemann-Franz law. (b) The subtracted additional negative contributions from charge neutral excitations using $\kappa^n_{xy}/T=\protect\kappa_{xy}/T-L_0\sigma_{xy}$. The dashed lines are the fit using the exponential function. (c) Collapse of all subtracted data after a proper rescaling with the fitting parameters, showing an excellent universal temperature dependence for all samples. The dashed line is the exponential function $e^{-T/T_0}$. The arrow marks the position of $T_0$ below which the data seem to deviate from the scaling. The inset shows the values of the derived $T_0$ as a function of doping. (d) $\protect\kappa_{xy}/TH$ in LSCO ($p=0.06$) for the magnetic field from 3 to 15 T. The hole contribution is negligible. The dashed lines are the exponential fit for each curve. The derived $T_0$ is shown in the inset as a function of the field.}
\label{fig1}
\end{figure}

We start with the thermal Hall conductivity discovered recently in both undoped and underdoped cuprates. Figure \ref{fig1}(a) reproduces the experimental data of $\kappa_{xy}/T$ measured in several compounds \cite{Grissonnanche2019Nature}. For La$_{2}$CuO$_{4}$ and lightly doped La$_{2-x}$Sr$_{x}$CuO$_{4}$ (LSCO, $p=0.06$) and La$_{1.8-x}$Eu$_{0.2}$Sr$_{x}$CuO$_{4}$ (Eu-LSCO, $p=0.08$), the data show large negative $\kappa_{xy}/T$ at low temperatures but approach zero above 100 K. Since the contribution of charge carriers ($\kappa^{e}_{xy}/T$) is negligible, the large negative thermal Hall conductivity must arise from additional charge neutral excitations ($\kappa^{n}_{xy}/T$). For La$_{1.6-x}$Nd$_{0.4}$Sr$_{x}$CuO$_{4}$ (Nd-LSCO) and Bi$_{2}$Sr$_{2-x}$La$_{x}$CuO$_{6+\delta }$ (Bi2201) with $p=0.2$, $\kappa_{xy}/T$ turns positive at high temperature and follows roughly the behavior of $\kappa^{e}_{xy}/T=L_{0}\sigma _{xy}$, where $L_{0}$ is the Lorenz number and $\sigma _{xy}$ is the Hall conductivity of charge carriers. Thus, the high temperature thermal Hall conductivity is dominated by charge carriers, while the deviation from the Wiedemann-Franz law at low temperatures reflects the contribution of charge neutral excitations. We have therefore a two-component formula:
\begin{equation}
\kappa_{xy}/T=\kappa^{n}_{xy}/T+\kappa^{e}_{xy}/T.
\end{equation}
Beyond the critical doping as in Nd-LSCO ($p=0.24$), the deviation from the Wiedemann-Franz law becomes indiscernible in experiment. Hence, $\kappa^{n}_{xy}/T$ is presumably associated with the pseudogap or the spin liquid. Its temperature dependence is shown in Fig. \ref{fig1}(b) after subtracting the charge contribution from the data.

Following the proposal in Ref. \cite{Yang2020PRL}, the thermal Hall conductivity of charge neutral excitations \cite{Xiao2010RMP,Matsumoto2011PRL,Qin2012PRB,Saito2019PRL} may follow an exponential scaling,
\begin{equation}
\frac{\kappa^n_{xy}}{TH}=\int d\epsilon \left(-\frac{\partial n}{\partial \epsilon }\right) \left(\frac{\epsilon-\mu}{T}\right)^{2}\mathcal{B}(\epsilon) \sim  e^{-T/T_{0}},
\label{THequation}
\end{equation}
where $\mathcal{B}(\epsilon)$ is their intrinsic or induced Berry curvature density under linear-in-field approximation \cite{Yang2020PRL}, $\mu$ is the chemical potential, and $n(\epsilon)$ denotes their corresponding Fermi-Dirac or Bose-Einstein distribution functions. It has been shown that the exponential scaling is robust for quite general cases and covers a wide intermediate temperature range \cite{Yang2020PRL}. The scaling occurs when the temperature is the order of the effective bandwidth $D$ of the Berry curvature density, and reflects a particular functional property bridging the low and high temperature limits. The characteristic temperature $T_0$ is typically a fraction of $D$. Indeed, as plotted in Fig. \ref{fig1}(c), the subtracted $\kappa^{n}_{xy}/T$ can all be scaled on a single curve following exactly the proposed scaling. The universal behavior is valid even for large field, as confirmed in Fig. \ref{fig1}(d) for the experimental data of LSCO ($p=0.06$) up to 15 T. As shown in the insets of Figs.~\ref{fig1}(c) and \ref{fig1}(d), the derived temperature scale $T_{0}$ is the order of 10 K in these compounds. Its magnitude decreases tentatively with increasing doping but is not very sensitive with the magnetic field. Later experiments suggest that the large $\kappa^{n}_{xy}/T$ might actually arise from chiral phonons through interaction with spins \cite{Grissonnanche2020NP}. In any case, the spin liquid seems to involve or at least induce a finite Berry curvature density to other carriers. The exponential scaling has already been confirmed in other cuprates or candidate spin liquid materials \cite{Grissonnanche2020NP,Boulanger2020NC,Yang2022PRB}.

Three lessons may be learned from the above thermal Hall analysis. First, there seem to exist multiple components like the spin liquid and the holes below the critical doping \cite{Barzykin2009AP}. Second, either the spin liquid itself exhibits fractionalized excitations, or there must be some way to bring topological contributions in cuprates \cite{Kaul2007NP}. Third, physical properties arising from Berry curvatures in a finite energy window may quite generally contain an exponential scaling term over a wide intermediate temperature range irrespective of model details \cite{Yang2020PRL}.

Hence, interpretation of experimental data may involve the combined effect of multiple excitations and their interactions. Earlier evidence for the existence of multiple carriers comes from the Knight shift measurements on different probing nuclei \cite{Hasse2008JPCM}, which has led to the proposal of a two-component scenario for describing  physical properties of underdoped cuprates \cite{Barzykin2006PRL,Barzykin2009AP}. It contains a spin liquid described by the two-dimensional Heisenberg model of localized Cu spins with a renormalized spin interaction, and a non-Landau Fermi liquid that contributes a temperature-independent constant to the magnetic susceptibility and the Knight shift. The magnetic properties are then governed mainly by the spin liquid \cite{Johnston1989PRL} and exhibit one parameter scaling for most of the temperature domain \cite{Nakano1994PRB,Wuyts1996PRB,Curro2005MRS}. Our analysis of the thermal Hall conductivity provides an example where contributions of both carriers show distinctive temperature dependence, thus preventing the one parameter scaling.

On the other hand, the electric transport properties are dominated only by charge carriers. They often exhibit one parameter scaling \cite{Luo2008PRB} but their interpretations are mostly uncertain. The $c$-axis resistivity obeys the scaling function, $\rho_c(T)\sim \frac{T}{\Delta}e^{\frac{\Delta}{T}}$, where $\Delta$ is the pseudogap, possibly contributed mainly by quasiparticles near the antinodal points owing to the interplay of anisotropic interplay hopping integral and the pseudogap effect \cite{Su2006PRB}. The in-plane resistivity $\rho$ exhibits a universal linear behavior at high temperatures in underdoped and optimally doped compounds \cite{Batlogg1994PhysC}. This strange metal behavior could result from scattering of nodal quasiparticles by the spin liquid, but it resists a good understanding \cite{Varma1989PRL,Nagaosa1990PRL,Ioffe1990PRL,Moriya1990JPSJ,Phillips2022Science}. 

By contrast, the Hall coefficient also exhibits one parameter scaling but lacks an explicit mathematical expression \cite{Hwang1994PRL}. A potential explanation of its high temperature behavior was given by the activation model for undoped and lightly doped LSCO \cite{Gorkov2006PRL,Ono2007PRB}, where electrons are assumed to be thermally excited across the charge transfer gap from the O valence band to the Cu upper Hubbard band at high temperatures, causing a positive normal Hall coefficient due to the larger mobility of the O holes. The model gives a good fit to the high temperature data in undoped and lightly doped LSCO but fails below 300 K or at larger doping, where the doped holes become dominant. Moreover, it predicts substantially larger carrier densities than anticipated and much smaller excitation gap than that observed in optical measurements. Thus, we still lack a good understanding of the Hall coefficient for most temperature and doping regions. Moreover, the activation model involves two types of contributions governed by different energy scales and is not in line with the idea of one parameter scaling found in optimal and overdoped LSCO \cite{Hwang1994PRL,Luo2008PRB}. It is therefore important to explore the scaling function of the Hall coefficient, which is likely to originate from some intrinsic contribution of doped holes.

If the above argument is correct, the Hall coefficient should contain a term from doped holes like \cite{Nagaosa2010RMP,Zhang2016NJP}
\begin{equation}
R_H=\rho^2\sigma_{xy}/H=\int d\epsilon\left(-\frac{\partial n_F}{\partial \epsilon }\right) \rho^2\mathcal{B}_h(T, \epsilon),
\label{RHequation}
\end{equation}
where $\mathcal{B}_h(T, \epsilon)$ is the Berry curvature density of the hole carriers \cite{Yang2020PRL}. Since the holes are doped on O sites and interact strongly with the Cu spins \cite{ZR1988PRB}, one may naively expect that they may also inherit some topological properties of the spin liquid. We have explicitly included the temperature in $\mathcal{B}_h(T, \epsilon)$ to emphasize that such an effect may be suppressed as the temperature is increased and the topological properties of the spin liquid diminish because of thermal fluctuations. 

At first glance, the above equation cannot lead to any useful conclusion, because both $\rho(T)$ and $\mathcal{B}_h(T, \epsilon)$ depend on model details and lack a good theoretical understanding. However, motivated by the success of the two-component analysis and the universal scaling in the thermal Hall conductivity, we explore whether or not similar phenomenological analysis may be applied to the Hall coefficient. Following Ref. \cite{Yang2020PRL}, it is easy to verify that the same conclusion still holds for Eq. (\ref{RHequation}) by assuming some simple analytic form of $\rho^2\mathcal{B}_h(T,\epsilon)$ \cite{note}. The proposed exponential scaling is a generic and robust property of the energy integral in the intermediate temperature range around $T\sim D_h$ so long as the hole carriers have a nonzero Berry curvature density in the finite energy window $D_h$. It is therefore not unreasonable to propose a similar exponential temperature dependence for the Hall coefficient, $R_H \sim \exp(-T/T_H)$. The scaling is independent of model details and bridges the low and high temperature limits, where the exponential scaling breaks down and the behaviors of the Hall coefficient depend on some microscopic details.

\begin{figure}[t]
\centerline{{\includegraphics[width=.5\textwidth]{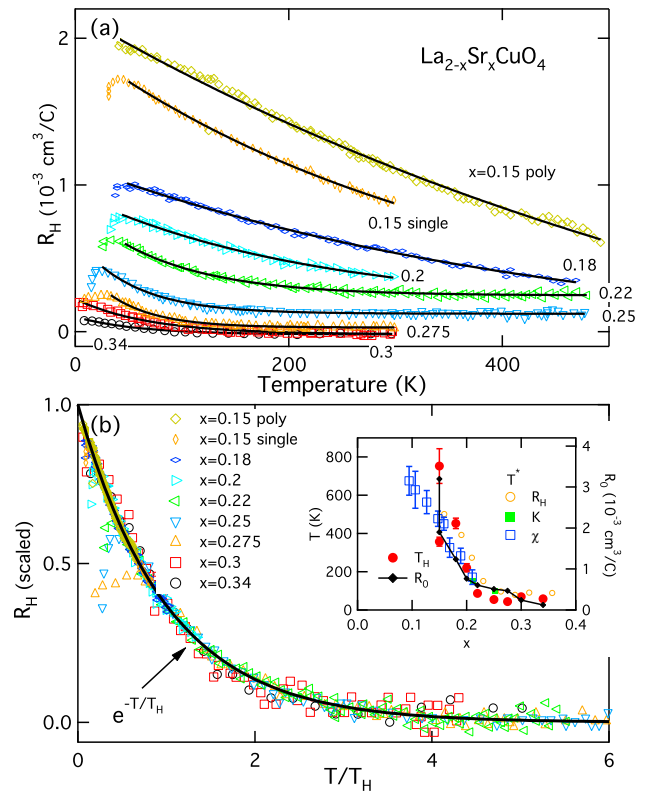}}}
\caption{(a) Hall coefficient of La$_{2-x}$Sr$_{x}$CuO$_{4}$ by Hwang {\sl et al.} reproduced from experiment \cite{Hwang1994PRL}. The solid lines are the fit using the scaling function, $R_H=R_\infty + R_0 e^{-T/T_H}$. (b) Collapse of all data after subtracting $R_\infty$ and rescaled by $T_H$ and $R_0$. The inset shows the derived values of $T_H$ and $R_0$ compared with those estimated previously by Hwang {\sl et al.} and those from the susceptibility and the Knight shift \cite{Hwang1994PRL}.}
\label{fig2}
\end{figure}

\begin{figure}[t]
\centerline{{\includegraphics[width=.5\textwidth]{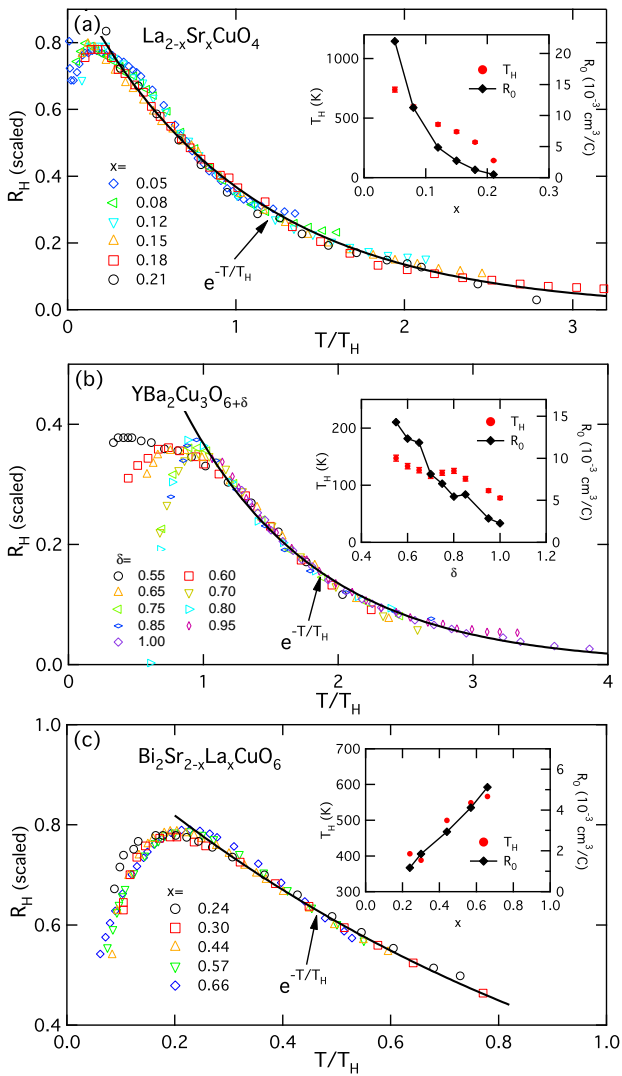}}}
\caption{Hall coefficient of (a) La$_{2-x}$Sr$_{x}$CuO$_{4}$ \cite{Ono2007PRB}, (b) YBa$_2$Cu$_3$O$_{6+\delta}$ \cite{Segawa2004PRB}, and (c) Bi$_{2}$Sr$_{2-x}$La$_{x}$CuO$_{6+\delta }$ \cite{Ando1999PRB}, reproduced from experiments. The data are collapsed for each compound according to Ref. \cite{Luo2008PRB}. The solid lines are the fit using the exponential scaling function. The derived values of $T_H$ and $R_0$ are given in the insets.}
\label{fig3}
\end{figure}

We now examine its validity in experiments. Figure \ref{fig2}(a) reproduces the Hall data of LSCO by Hwang {\sl et al.} for $0.15\le x\le 0.34$ \cite{Hwang1994PRL}. Applying the scaling function, 
\begin{equation}
R_H=R_\infty + R_0 e^{-T/T_H},
\label{RHscaling}
\end{equation}
in which $R_\infty$, $R_0$, and $T_H$ are free parameters independent of the temperature, we find an excellent fit (solid lines) to the data over a wide temperature range for all these doping levels. The data are then rescaled using the fitting parameters and replotted in Fig. \ref{fig2}(b). We see a good collapse on the exponential function. The scaling only breaks down at very low temperatures, where a maximum occurs for each doping. The derived values of the parameter $T_H$ are given in the inset and compared with those determined previously by Hwang {\sl et al.} and from the magnetic susceptibility and Knight shift scaling  \cite{Hwang1994PRL}. There is an overall agreement in their doping dependence. For $x>0.2$, our derived $T_H$ is smaller, which is not unexpected since it is presumably only a fraction of $D_h$. The consistency suggests that the Hall coefficient might indeed have a nontrivial origin associated with the spin liquid or the pseudogap for $x\le 0.22$. Interestingly, its magnitude $R_0$ also follows a similar doping dependence, possibly reflecting the reduced strength of the spin liquid with increasing hole density. For $x\ge 0.25$, $T_H$ is almost doping independent, while $R_0$ is extrapolated to zero for $x>0.34$.  The fact that the scaling is still satisfied suggests the possibility of hole density inhomogeneity in this region but awaits further clarification. Note that the large magnitude of $T_H$ compared to $T_0$ from the thermal Hall conductivity may reflect the very different coupling strengths of the O holes and the phonons with the Cu spins, which should be taken into consideration in a microscopic theory.

To further support our theory, we apply the scaling function to other measurements including LSCO by Ono {\sl et al.} \cite{Ono2007PRB}, YBa$_2$Cu$_3$O$_{6+\delta}$ \cite{Segawa2004PRB}, and Bi$_{2}$Sr$_{2-x}$La$_{x}$CuO$_{6+\delta }$ \cite{Ando1999PRB}. One parameter scaling has been confirmed for all these data, but a common scaling function was not found for three materials \cite{Luo2008PRB}. Figure \ref{fig3} reproduces the experimental data and their collapse for each compound as proposed in \cite{Luo2008PRB}. Again, we find excellent fit (solid lines) for all three compounds. There does exist a common scaling function, despite their very different temperature ranges and low temperature behaviors that make it difficult to find without a proper theoretical guidance. For LSCO, the scaling extends to underdoped samples and applies over the wide range of $0.05\le x\le 0.34$. For each doping, the scaling breaks down at some low or high temperatures. However, the collapsed data show an overall tendency that agrees well with the scaling function over a much broader temperature range, suggesting that our theory indeed captures some intrinsic properties common for all doped cuprates. The derived parameters $T_H$ and $R_0$ for all three materials are plotted in the insets and, again, show roughly similar tendency with doping. 

\begin{figure}[t]
\centerline{{\includegraphics[width=.45\textwidth]{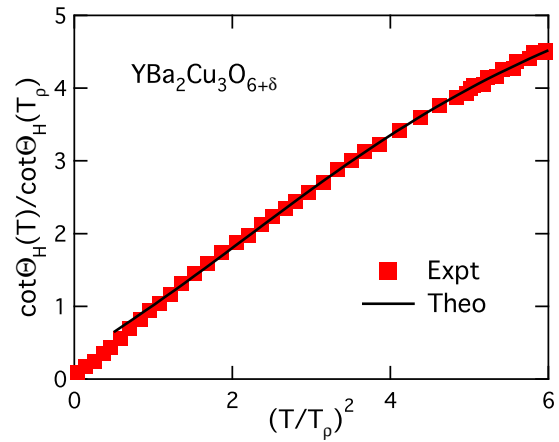}}}
\caption{Comparison of our predicted scaling (solid line) with the collapsed Hall angle data for dozens of YBa$_2$Cu$_3$O$_{6+\delta}$ samples reproduced from Ref. \cite{Wuyts1996PRB}, where $T_\rho$ is some temperature scale dervied from resistivity scaling in the original literature.}
\label{fig4}
\end{figure}

Compared to the activation model, our theory covers a much wider temperature and doping range and is consistent with the idea of one parameter scaling that has long been established for the Hall coefficient \cite{Hwang1994PRL}. We cannot exclude the presence of some thermally activated carriers at very high temperatures, which might also interact with the spins. Although a microscopic understanding is not available at this moment, the simplicity of our scaling function and its excellent agreement with experimental data make it a favorable scenario completely different from the activation model. While the latter is governed by the normal Hall coefficient of thermally excited carriers across the charge transfer gap, our theory suggests that the Hall coefficient is associated with nonzero Berry curvature density of doped holes. It is crucial if this can be verified in other probes.

Actually, one immediate prediction of our theory is that, following the same argument \cite{note}, the Hall angle should also exhibit the exponential scaling in some temperature range, namely, $\tan\Theta_H=\rho\sigma_{xy}=A+Be^{-T/T_\theta}$, where $T_\theta$ could be slightly different from $T_H$ \cite{note}. We examine this prediction with the data in YBa$_2$Cu$_3$O$_{6+\delta}$, which experimentally show good collapse for dozens of samples \cite{Wuyts1996PRB}. As plotted in Fig. \ref{fig4}, the overall tendency of their collapsed curve indeed obeys our scaling function over a wide high temperature range, in sharp contrast to the famous $T^2$ law proposed by Anderson \cite{Anderson1991PRL}, which is only valid in a very limited low temperature window $0.3\le (T/T_\rho)^2\le 1.5$ \cite{Wuyts1996PRB}.

Question remains on how our theory could be realized microscopically. It is well possible that some other mechanism is responsible for the exponential scaling. However, systematic comparisons of the one parameter scaling for the $c$-axis resistivity, the in-plane resistivity, the Hall coefficient, the magnetic susceptibility, the Knight shift, the spin-lattice relaxation rate, and the Seebeck coefficient suggest that they are all governed by the same energy scale associated with the pseudogap in the underdoped region \cite{Luo2008PRB,Barzykin2009AP}. In the two-component scenario, it is further determined by the effective spin interaction of the spin liquid, which decreases with increasing hole doping and diminishes beyond the critical doping \cite{Barzykin2009AP}. The same energy scale seems to govern all different types of responses of the spin and charge degrees of freedom. Our proposed scaling for the Hall coefficient is consistent with this overall picture and supports a unified mechanism for the basic physics of cuprates.

This work was supported by the National Key R\&D Program of China (Grant No. 2022YFA1402203), the National Natural Science Foundation of China (Grants No. 12174429, No. 11974397),  and the Strategic Priority Research Program of the Chinese Academy of Sciences (Grant No. XDB33010100).

\end{document}